\documentclass{aastex631}

\usepackage{amsmath}

\begin{document}

\title{Internal activities in a solar filament and heating to its threads}

\correspondingauthor{Zhenghua Huang}
\email{z.huang@sdu.edu.cn}

\author[0000-0002-0210-6365]{Hengyuan Wei}
\affiliation{Shandong Key Laboratory of Optical Astronomy and Solar-Terrestrial Environment, Institute of Space Sciences, Shandong University, Weihai 264209, Shandong, China}

\author[0000-0002-2358-5377]{Zhenghua Huang}
\affiliation{Shandong Key Laboratory of Optical Astronomy and Solar-Terrestrial Environment, Institute of Space Sciences, Shandong University, Weihai 264209, Shandong, China}

\author[0000-0001-7693-4908]{Chuan Li}
\affiliation{School of Astronomy and Space Science, Nanjing University, Nanjing 210023, China}
\affiliation{Key Laboratory for Modern Astronomy and Astrophysics (Nanjing University), Ministry of Education, Nanjing 210023, China}

\author[0000-0003-4804-5673]{Zhenyong Hou}
\affiliation{School of Earth and Space Sciences, Peking University, Beijing, 100871, People's Republic of China}

\author[0000-0002-1190-0173]{Ye Qiu}
\affiliation{School of Astronomy and Space Science, Nanjing University, Nanjing 210023, China}
\affiliation{Key Laboratory for Modern Astronomy and Astrophysics (Nanjing University), Ministry of Education, Nanjing 210023, China}

\author[0000-0002-8827-9311]{Hui Fu}
\affiliation{Shandong Key Laboratory of Optical Astronomy and Solar-Terrestrial Environment, Institute of Space Sciences, Shandong University, Weihai 264209, Shandong, China}

\author[0000-0003-2686-9153]{Xianyong Bai}
\affiliation{National Astronomical Observatories, Chinese Academy of Sciences, Beijing, 100012, People's Republic of China}

\author[0000-0001-8938-1038]{Lidong Xia}
\affiliation{Shandong Key Laboratory of Optical Astronomy and Solar-Terrestrial Environment, Institute of Space Sciences, Shandong University, Weihai 264209, Shandong, China}



\begin{abstract}

Filaments are one of the most common features in the solar atmosphere, and are of significance in solar, stellar and laboratory plasma physics.
Using data from the Chinese H$\alpha$ Solar Explorer, the Solar Upper Transition Region Imager and the Solar Dynamics Observatory, we report on multiwavelength imaging and spectral observations of the activation of a small filament.
The filament activation produces several localized dynamic brightenings, which are probably produced by internal reconnections of the braided magnetic fields in the filament.
%a filament which is around 71 arcsec and interaction between the filament threads and ambient magnetic field.
%The eruption of filament drives several localized dynamic brightenings.
%The localized brightenings are the results of internal reconnections of the filament possibly resulted from the braided magnetic field.
The filament expands during the activation and its threads reconnect with the ambient magnetic fields, which leads to the formation of hot arcades or loops overlying the filament.
The thermal energy of each of these localized brightenings is estimated in the order of $10^{25}-10^{27} erg$ and the total energy is estimated to be $\sim1.77 \times 10^{28} erg$.
Our observations demonstrate that the internal magnetic reconnections in the filament can lead to localized heating to the filament threads and prompt external reconnections with ambient corona structures, and thus could contribute to the energy and mass transferring into the corona.

\end{abstract}



\section{Introduction} \label{sec:intro}

\par
Filaments, also called prominences while appearing at the limb, are common features in the solar atmosphere with dense, cool and partially ionized plasma.
They anchor to the photosphere with anti-polarity, extend along the polarity inversion lines (PILs) and outward to the hot corona.
They could be unambiguously observed in chromospheric lines such as H$\alpha$ and EUV passbands such as 304\,\AA\, and 171\,\AA\, \citep{2014LRSP...11....1P}.
The instabilities and eruptions of filaments could lead to solar flares and/or coronal mass ejections \citep{2011LRSP....8....1C,2011LRSP....8....6S,2012ApJ...750...12S,2017LRSP...14....2B,2019ApJ...885L..11S}.
Thanks to the high-resolution facilities, more and more fine structures are reported, which show complex dynamics in the filament and raise much more puzzles about the phenomenon \citep{2018LRSP...15....7G}.

\par
Filaments are formed in filament channels along the PILs \citep{1998SoPh..182..107M,2010SSRv..151..333M} and the chromospheric filamentary threads are aligned with the filament channels, whose magnetic structures are thought to be sheared-arcades or twisted flux-ropes \citep{1999ApJ...518L..57A,2000ApJ...539..954D,2004ApJ...609.1123F,2006JGRA..11112103G,2002ApJ...567L..97A}.
Many studies have demonstrated that numbers of eruptive filament events are associated with internal reconnection and thus internal reconnection plays an important role in filament associated activities.
\citet{2001ApJ...552..833M} observed magnetic explosion events with flares and coronal mass ejections (CMEs), and the internal reconnection of filaments is crucial in the onset and growth of these eruptive events.
\citet{2004ApJ...613.1221S} reported that the external and internal reconnections cause the eruption of the filament-carrying magnetic cavity.
\citet{2012ApJ...745..164S} reported that the external and internal reconnection of filament could produce simultaneously bubble-like and jet-like CME.
\citet{2004ApJ...611..545G} illustrated that the external reconnection leads to the internal reconnection of filament, which releases the helix and heats a two-ribbon flare.
\citet{2009A&A...498..295T} found that most partial eruptions of prominences could be explained by partially-expelled-flux-rope model which is associated with internal reconnection.
\citet{2018ApJ...856...48C} suggested that the splitting of filament involving internal reconnection could induce eruptions of filament.
\citet{2018ApJ...853L..26H} suggested that the free magnetic energy could be released into upper atmosphere through reconnection process.
Many observations and simulations indicated that the internal reconnection of filaments or mini-filaments could trigger jets in the corona\,\citep{2015Natur.523..437S,2016ApJ...832L...7P,2017Natur.544..452W,2017ApJ...851...67S,2019ApJ...883..104S,2022ApJ...936...51W}.

\par
Many complex motions, small-scale filament-eruption-like features and fine structures in filament are also widely reported in previous literatures.
\citet{2015ApJ...814L..17S} indicated that the magnetic reconnection at separator could cause upflows and this could intrude into downflow area and form counter-streaming mass flows.
\citet{2020ApJ...889..187S} reported several small-scale eruptions of  sub-minifilaments which could be caused by magnetic reconnections or acoustic waves.
\citet{2023A&A...673A..83L} studied the event of mini-filament and found that the mini-filament eruption experiences both internal and external reconnection, which transfers mass and flux to ambient corona.
\citet{2019ApJ...887...56T} found brightenings at the ends of loops in an arch filament system, which lie on the opposite-polarity magnetic convergence, and these brightenings are likely caused by magnetic cancellation.
\citet{2020ApJ...899...19C} found a series of coronal mini-jets in an activated tornado-like prominence, and these mini-jets were thought to be the results of fine-scale external or internal reconnections between the prominence magnetic field and background field or the twisted or braided fields in the prominence.
\citet{2022A&A...663A.127L} demonstrated that the repeated, intermittent small-scale H$\alpha$/EUV bursts at the same place were related to the growth of the filament channel.
\citet{2019ApJ...874..176H} reported several localized brightenings in filament in microwaves with a temporal variations of 3-5 s, which correspond to the EUV bright threads, fibers or spots, and these brightenings might be caused by internal reconnection or electric field acceleration.
%first appear at the top of the filament and scatter in the whole filament subsequently.
\citet{2016SoPh..291.2373Z} reported some brightenings in the lengthening and widening process of filament, these brightenings form at the interface between different short and thin threads and are the results of reconnection among those threads.

\par
In this paper, we report on multiwavelength imaging and spectroscopic observations of the activation of a filament.
In this case, the activated filament is confined with a series of localized small-scale brightenings, and results in over-lying hot arcades.
We will investigate in-depth how these small-scale localized brightenings form and the thermal and dynamic evolution of the filament.
In what follows, we give the data description in Section \ref{sec:obs}, results in Section \ref{sec:res}, discussion in Section \ref{sec:dis} and conclusions in Section \ref{sec:con}.

\section{Observations} \label{sec:obs}
The data were observed on 2022 November 1 from 16:50\,UT to 17:15\,UT with a target of AR 13135, by the space-borne Atmospheric Imaging Assembly \citep[AIA,][]{2012SoPh..275...17L} and the Helioseismic and Magnetic Imager \citep[HMI,][]{2012SoPh..275..207S} onboard the Solar Dynamics Observatory \citep[SDO,][]{2012SoPh..275....3P}, the H$\alpha$ Imaging Spectrograph\,\citep[HIS,][]{2022SCPMA..6589605L} onboard the Chinese H$\alpha$ Solar Explorer \citep[CHASE,][]{2022SCPMA..6589602L,2022SCPMA..6589604Z} and the Solar Upper Transition Region Imager \citep[SUTRI,][] {2023RAA....23f5014B,2023RAA....23i5009W}.

The data analyzed involve the line-of-sight magnetograms observed by HMI, images taken by the AIA EUV passbands of 94, 131, 171, 193, 211, 304 and 335\,\AA\,, images with SUTRI at Ne \uppercase\expandafter{\romannumeral7} 465\,\AA\, and the spectral data of H$\alpha$ and Fe \uppercase\expandafter{\romannumeral1} lines with HIS.

\par
The AIA data  have a pixel size of 0.6\arcsec with a cadence of 12 s for EUV images and 24 s for 1600\,\AA\, images.
The HMI data have a pixel size of 0.6\arcsec and a cadence of 45 s.
Both AIA and HMI data are reduced by the standard procedure of {\it aia\_prep.pro} in solarsoft.

\par
The HIS was running in a ``Raster Scanning Mode'' (RSM) to obtain spectra of the full solar disk.
%HIS has an exposure time of about 5\,ms, and can obtain spectral data of the full disk in about 50\,s.
A full-disk scanning takes about 60 seconds.
The present dataset includes 10 rasters of the full-disk obtained from 16:45:04\,UT to 16:55:42\,UT.
The spatial resolution of the binning mode data used in this paper is 1.04\arcsec/pixel, while the spectral resolution is 0.048 \AA/pixel.
We analyze the spectral data of H$\alpha$ (6559.7 - 6565.9 \AA) and Fe \uppercase\expandafter{\romannumeral1} (6567.8 - 6570.6 \AA) in this study.
The latter one is applied to co-align with the AIA and HMI observations.
The production of spectral data is level 1 after full calibrations\,\citep{2022SCPMA..6589603Q}.
%, and do not need further calibration.2022SCPMA..6589606Z

\par
The SUTRI images have a spatial resolution of 1.23\arcsec/pixel and 30 s cadence.
It observes the Sun at Ne \uppercase\expandafter{\romannumeral7} 465\,\AA\, whose formation temperature is around 0.5 MK \citep{2017RAA....17..110T}.
The data production downloaded is high-level scientific data which have been fully calibrated by the instrument team.

\par
The images from different passbands and instruments have been aligned by the reference structures (such as sunspots, filaments and networks) in the passbands with the closest representative temperatures.

\begin{figure*}
\includegraphics[trim=0cm 1.7cm 0cm 2.5cm,width=\textwidth]{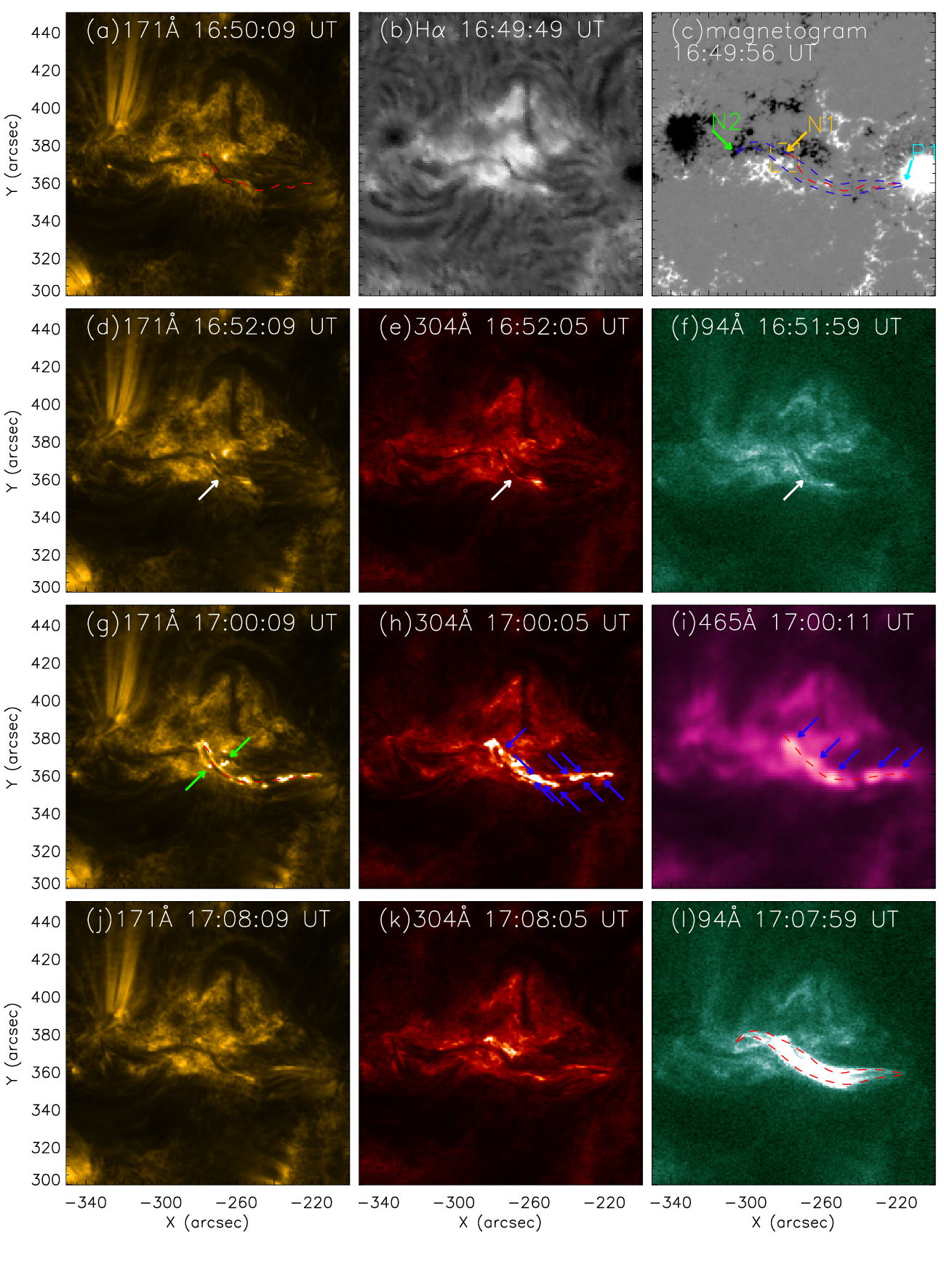}
\caption{Overview of the region of interest in AIA 171\,\AA\,, 304\,\AA\,, 94\,\AA\,, SUTRI 465\,\AA\,, HIS $H \alpha$ and HMI line-of-sight magnetogram.
(a): The red dash line outlines the shape of the filament.
(c): The cyan arrow marked as P1 points to the positive polarity of the filament and arcade A2, the yellow arrow marked as N1 points to the negative polarity of the filament and the green arrow marked as N2 points to the negative polarity of the arcade A2.
The red dash line resembles the trajectory of the filament and the blue dash lines resemble the trajectory of the arcade A2.
%The yellow dash rectangle shows the field of view of the Figure \ref{fig:hmi}.
(d)-(f): The white arrows point to the bright arcade A1 overlying the filament.
%The blue lines in panel (e) show the contour of blue shifts at the level of 5 km/s.
(g): The green arrows point to the brightenings associated with the formation of the arcade A2.
The red dash line shows the trajectory that obtains the slice-time plot.
(h)-(i): The blue arrows point to the localized brightenings.
The red dash line in (i) shows the trajectory that obtains the slice-time-plot.
(l): The red dash lines show the trajectory of the arcade A2.
An associated animation is available online.
The animation shows the evolution of this region in these passbands from 16:50:11\,UT to 17:49:47\,UT.
The real-time duration of the animation is 30 s.
\label{fig:ove}}
\end{figure*}

\section{Results} \label{sec:res}
The event occurs on the disk from 16:50 to 17:15\,UT and the overview of this event is shown in Figure \ref{fig:ove}.
A filament with a length of around 71 arcsec can be unambiguously observed in HIS H$\alpha$ passband and AIA 304 \,\AA\, and 171 \,\AA\, passbands (see the red dash lines in panel a).
This filament extends along the PIL and connects the positive polarity P1 and negative polarity N1 (see panel c), which indicates that the magnetic configuration of the filament is a normal-polarity topology \citep{2014ApJ...784...50C}.
At around 16:51:29\,UT, the filament activation begins to occur, with the brightening of an arcade (hereafter A1) overlying the filament (see white arrows in panels d-f).
After that the filament destabilizes and starts to activate, several localized brightenings appear in most AIA EUV passbands and SUTRI 465\,\AA\,passband (see the blue arrows in panels h and i).
In addition, during the process of activation, the filament expands obviously, leading to the appearence of two brightenings at both sides of the filament at around 17:00:11\,UT, with an unidirectional flow towards east end of the filament appearing (see brightenings pointed by green arrows in panel g).
It is followed by the formation of hot arcades overlying the filament (hereafter called A2), which can be only observed in AIA 94\,\AA\, passband and connect the positive polarity P1 and negative polarity N2 (see the blue dash lines  in panel c and red dash lines in panel l).
%During the eruption process, the filament expands obviously.
%Using the H $\alpha$ spectral data from CHASE, we find a blue shift region at the east footpoint of the loop which enhances during the eruption, and this indicates there exists an enhanced outflow at the footpoint (see the blue contour in panel (e) in Figure \ref{fig:ove} and the white contours in Figure \ref{fig:dop}).

\begin{figure*}
\includegraphics[trim=0cm 0cm 0cm 0cm,width=\textwidth]{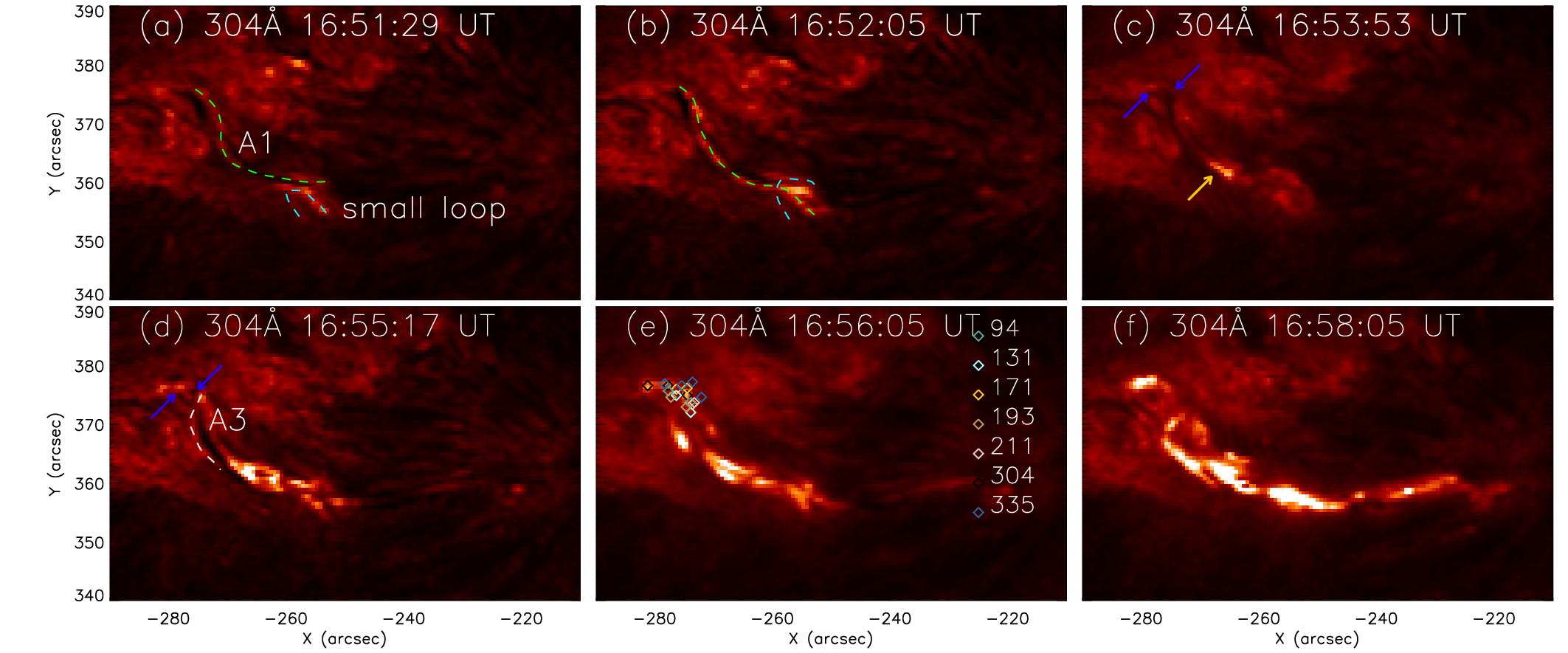}
\caption{The triggering process of the filament activation in AIA\,304\,\AA\,from 16:51:29\,UT to 16:58:05\,UT.
(a): The snapshot of AIA\,304\,\AA\, observation at 16:51:29\,UT, the green dash line marks the position of an arcade overlying the filament and the cyan dash line marks a nearby loop.
(b): The snapshot of AIA\,304\,\AA\,observation at 16:52:05\,UT, the green and cyan dash lines mark the arcade and loop after reconnection.
(c): The snapshot of AIA\,304\,\AA\,observation at 16:53:53\,UT, the blue arrows point to the eastern filament footpoint which begins to split. The yellow arrow points to the first localized brightening.
(d): The snapshot of AIA\,304\,\AA\,observation at 16:55:17\,UT, the blue arrows point to the eastern filament footpoint which has splitted, and the white dash line marks another brightened arcade.
(e): The snapshot of AIA\,304\,\AA\,observation at 16:56:05\,UT, the diamonds in different colours resemble the the position of bright points near the eastern footpoint of filament in AIA\,94, 131, 171, 193, 211, 304, 335\,passbands.
(f): The snapshot of AIA\,304\,\AA\,observation at 16:58:05\,UT.
\label{fig:tri}}
\end{figure*}

\par
\begin{figure*}
\includegraphics[trim=0cm 0cm 0cm 0cm,width=\textwidth]{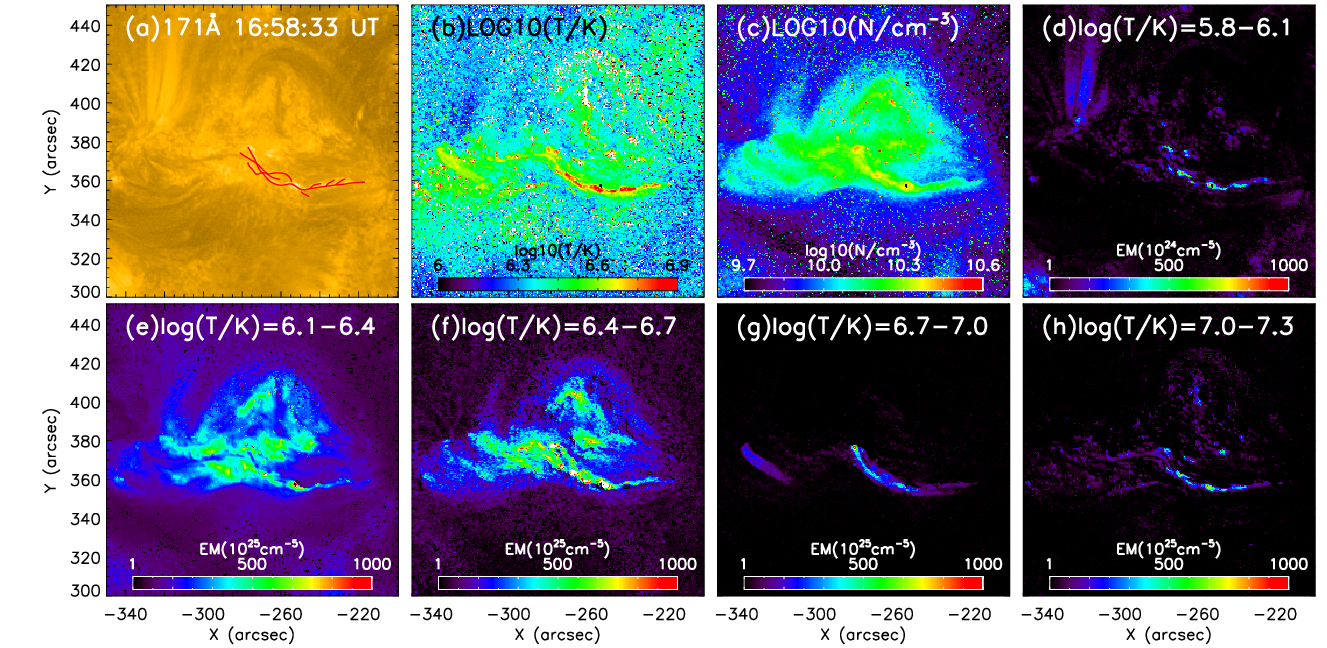}
\caption{Differential emission measure of the region of interest.
(a): The enhanced logarithm image with unsharp\_mask operator in AIA 171\,\AA\, in 16:58:53 UT.
The red lines mark the trajectory of the brightened threads of filament.
(b): The weighted temperature distribution of the region.
(c): The density distribution of the region.
(d)-(h): The emission measure of the region in the range of log(T/K) = 5.8-7.3.
An associated animation of this area is available online.
The animation includes the evolution of the region seen in AIA 171\,\AA, temperature distribution and density distribution from 16:49:59\,UT to 17:43:59\,UT.
The real-time duration of the animation is 30 s.
\label{fig:dem}}
\end{figure*}

Before the activation of the filament at 16:51:29\,UT, we observe the arcade A1 overlying the filament with a small loop nearby in the observation of AIA 304\,\AA\,passband (marked with the green and cyan dash lines in panel (a) in Figure \ref{fig:tri} respectively).
At 16:51:53\,UT, the interface between the A1 and small loop brightens, forming a new post-reconnected bright arcade overlying the filament and a post-reconnected small loop marked by green and cyan dash lines in panel (b) in Figure \ref{fig:tri}.
After this process, the first localized brightening appears at the place pointed by a yellow arrow in panel (c) in Figure \ref{fig:tri} while the filament threads near the east footpoint start to split (see the blue arrows in panel (c) in Figure \ref{fig:tri}).
At 16:55:17\,UT, another arcade (hereafter called A3) overlying the filament marked with white dash line in panel (d) in Figure \ref{fig:tri} brightens, and the filament threads near the east footpoint continue splitting during this period (blue arrows in panel (d) in Figure \ref{fig:tri}).
Later, more localized brightenings appear after destabilization of the filament.
With the boundary enhancement operator of {\it unsharp\_mask} in {\it ssw}, we are able to trace several bright threads of the filament in the enhanced AIA 171\,\AA\, passband images (see the red lines in Figure \ref{fig:dem} (a)).
We find that the localized brightenings are located on the threads and some of them are located at the cross section of these bright threads.
At 16:56:05\,UT, a series of brightenings appear near the east footpoint of the filament, and they seem to have different locations while seen on different AIA EUV passbands (see the diamonds with different colors in panel (e) in Figure \ref{fig:tri}).

\begin{figure*}
\includegraphics[trim=0cm 0cm 0cm 0cm,width=\textwidth]{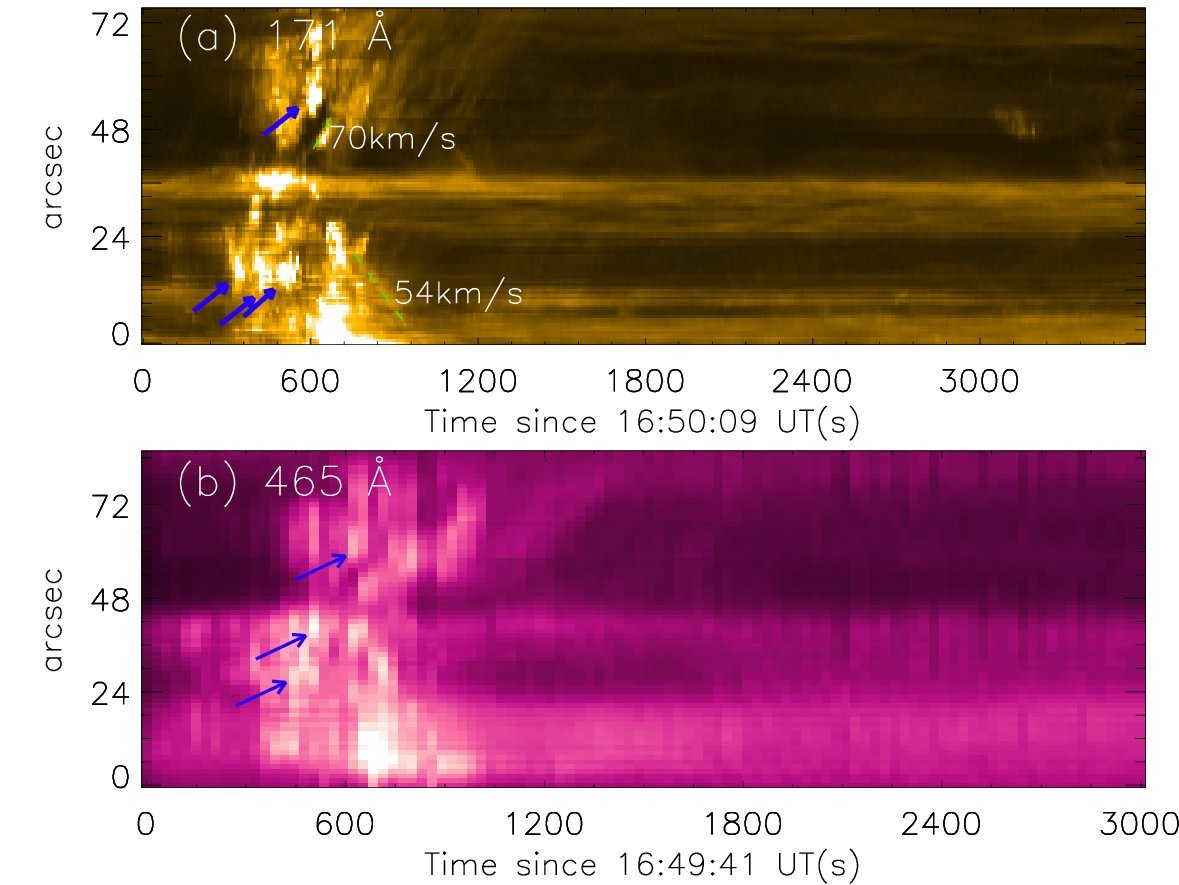}
\caption{Slice-time plot along the red dash lines in Figure \ref{fig:ove} (g) and (i).
The blue arrows in panel (a) and (b) point to several still localized brightenings.
The green dash lines in panel (a) trace some moving brightenings.
\label{fig:st}}
\end{figure*}

\par
To study the dynamics of the localized brightenings on the filament, we make the slice-time plot along a slit along the filament (see red dash lines in panels (g) and (i) in Figure \ref{fig:ove}).
In Figure \ref{fig:st}, we find that the dynamics of these localized brightenings shows similar trend in both AIA 171\,\AA\,passband and SUTRI 465\,\AA\,passband.
Some localized brightenings stay still (see the blue arrows in panel (a)-(b)) while some others propagate along the filament with speeds in the range of 50--70 km/s (see the green dash lines).
The stationary localized brightenings appear prior to those propagating localized brightenings.
Furthermore, some still localized brightenings occur at the same place repeatedly.
%The speeds of the moving brightenings derived from gradient of the bright trajectory are about 54 km/s and 70 km/s.

\begin{figure*}
\includegraphics[trim=0cm 0cm 0cm 0cm,width=\textwidth]{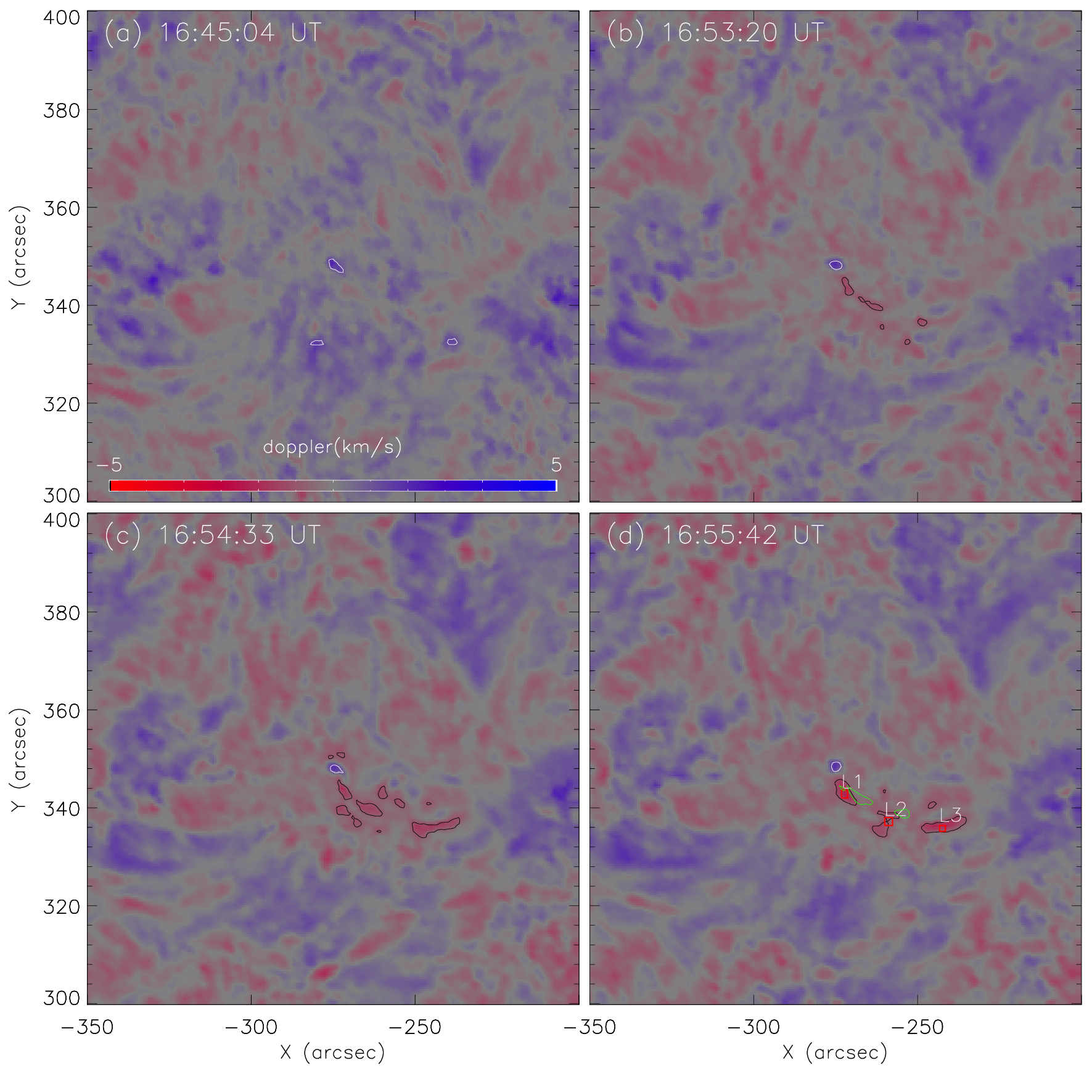}
\caption{Dopplergrams of the region of interest in the wavelength of $H \alpha$.
(a)-(d): The dopplergrams in 16:45:04 UT, 16:53:20 UT, 16:54:33 UT and 16:55:42 UT.
The white lines contour the blue shift at the level of 1 km/s.
The black lines in panel (b)-(d) show the contours of red shift at the level of 0.8 km/s.
The red rectangles marked with L1, L2, L3 show the positions where we obtain the spectral profiles shown in figure \ref{fig:ras}.
The green contours in panel (d) mark the localized brightenings in AIA 171\,\AA\, on 16:55:45\,UT at a level of 3000\,DN.
\label{fig:dop}}
\end{figure*}

%\begin{figure*}
%\includegraphics[trim=0cm 0cm 0cm 0cm,width=\textwidth]{fig_171_dop.eps}
%\caption{Observations in AIA\,171\,\AA\, and contours of blue and red shift at nearby time.
%(a)-(c): The observations in 16:53:21 UT, 16:54:33 UT and 16:55:45 UT.
%The white lines contour the blue shift at the level of 1 km/s.
%The green lines show the contours of red shift at the level of 0.8 km/s.
%\label{fig:171_dop}}
%\end{figure*}

\par
\begin{figure*}
\includegraphics[trim=0cm 0cm 0cm 0cm,width=\textwidth]{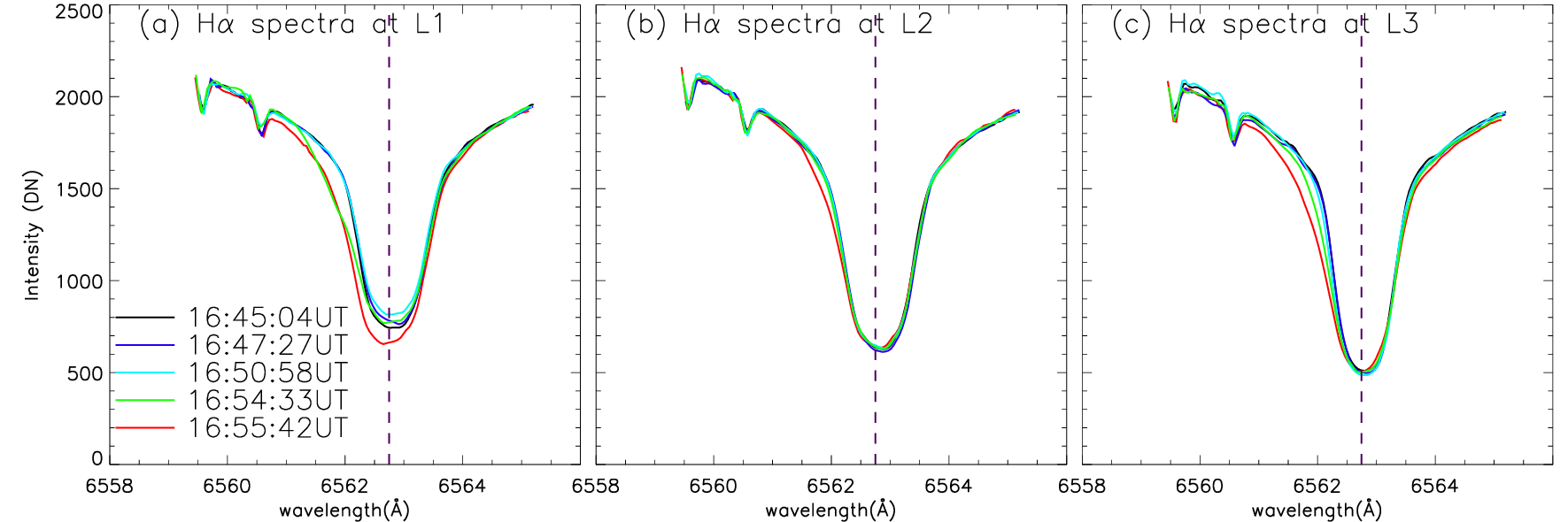}
\caption{Evolution of the spectral profiles at the red shift regions marked as L1, L2, L3 in Figure \ref{fig:dop}.
The black, blue, cyan, green and red lines in panels (a)-(c) show the spectral profiles at L1, L2 and L3 at 16:45:04\,UT, 16:47:27\,UT, 16:50:58\,UT, 16:54:33\,UT and 16:55:42\,UT.
The purple dash line represents the position of line core of reference spectral profile.
\label{fig:ras}}
\end{figure*}

Using the H$\alpha$ spectral data from CHASE, we obtain the dopplergram of the region of interest with the moment method\,\citep{2020ApJ...896..154Y}.
The reference spectrum is obtained from the average of those in the field-of-view shown in Figure\,\ref{fig:ove}(b).
At 16:47:27\,UT, we observe that some red shifted regions with speeds at around 1 km/s emerge at the places marked (see black contours in panel (b)-(d) in Figure \ref{fig:dop}) and the sizes of these red shifted regions seem to expand after 16:53:20\,UT, when the first localized brightening appears.
%Please note that CHASE does not have observations after 16:55:42\,UT.
These red shifted regions are located next to the localized brightenings in AIA EUV passbands (see the green contours in panel (d) in Figure \ref{fig:dop}).
We further investigate the evolution of the average spectral profiles in three red shifted structures marked with rectangles with L1, L2 and L3 in panel (d) of Figure \ref{fig:dop}.
We can see that the H$\alpha$ profiles in these regions are red-shifted in prior to the event (see the spectral profiles in Figure \ref{fig:ras} at 16:45:04\,UT and 16:47:27\,UT).
During the event, H$\alpha$ profiles show deeper blue wings and line centers while the red wings are almost not affected (see the spectral profiles in Figure \ref{fig:ras} at 16:54:33\,UT and 16:55:42\,UT), although the whole profiles are remaining red-shifted.
The red shift of H$\alpha$ centriod might suggest the sinking of the filament and internal energy release. While the blue-wing absorption indicates the injection of filament materials or the upward motion of filament materials.
%This indicates that the filament are expanding and/or rising during the event.
The Ti\,{\sc ii} 6559.5\,\AA\ and Si\,{\sc i} 6560.6\,\AA\ seem to be not affected by the event, except in the cases where the far blue wing of H$\alpha$ profile is strongly affected.

%After the east footpoint of the filament shifts to the blue shift region, the filament erupts with several localized brightenings (see the blue arrows in panel (h) and (i) in Figure \ref{fig:ove}), and the dopplergram shows some localized red shift regions on the filament (see the black contours in Figure \ref{fig:dop} (c)-(d)).
%There exists displacement between the location of localized red shift region and localized brightenings (see black contours in Figure \ref{fig:171_dop} which marks the position of red shift region).
%The first brightening appears at the middle of the filament and other brightenings scatter on the filament.
%These brightenings could also be observed by SUTRI though much more ambiguous.
%The position marked with white arrow in Figure \ref{fig:dop} (d) shows both red and blue shifts, which indicates rotation or relaxation at this place.
%However, the red or blue shift signatures in Fe \uppercase\expandafter{\romannumeral1} are not prominent.

\par
To investigate the thermal dynamics and density evolution of this region, we apply differential emission measure (DEM) analysis \citep{2015ApJ...807..143C,2018ApJ...856L..17S} based on the observations taken by AIA 94, 131, 171, 193, 211 and 335\,\AA\, passbands and get the emission measure (EM) of the region.
Using the EM data, we inverse the weight-average temperature and density distribution and evolution of this area with Equations \ref{equ:dem_inv},

\begin{equation}\label{equ:dem_inv}
\begin{aligned}
  EM = \int DEM(T)\, dT,\\
  \bar{T} = \frac{\int DEM(T) \times T\, dT}{\int DEM(T)\, dT},\\
  n = \sqrt{\frac{EM}{L}},
\end{aligned}
\end{equation}

where DEM is the differential emission measure, T is the temperature and L is the depth/width of the structures along line-of-sight\,\citep{2012ApJ...761...62C}.
Here, we use the width of filament to estimate the accurate density of filament.
Before activation of the filament, the arcade A1 overlying the filament shows a temperature at around log (T/K) = 6.6 and already shows a weak density enhancement.
%(see the white arrow in panel (d)-(f) in Figure \ref{fig:ove})
During the activation process, the temperature response first enhances at the middle part of the filament with the highest temperatures given at the localized brightenings, while the density increase starts from the east end of the filament and spreads toward the west (see the associated animation of Figure \ref{fig:dem}).
The localized brightenings appear in a wide range of temperature from log (T/K) = 5.5 to log (T/K) = 7.3 and the average temperature is estimated to be over $8 \times 10^6 \rm K$.
The filament is heated to near  $8 \times 10^6 \rm K$ during the activation process.
The arcades A2 overlying the filament, which can be only observed in AIA 94\,\AA\, passband, have a strong EM response in log (T/K) = 6.7-7.0 and are cooled down in tens of minutes after the activation, which is shown as the disappearance of bright structure in AIA 94\,\AA.
%(see the red dash lines in panel (l) in Figure \ref{fig:ove})
The average temperature of these arcades A2 is around log (T/K) = 6.7-6.8 (see associated animation of Figure \ref{fig:dem}).
%, which can be only observed in AIA 94\,\AA\,passband (see red dash lines in panel (l) in Figure \ref{fig:ove}),
%??It should be noticed that the DEM derived here only represents the prominence-corona transition region (PCTR), which is the foreground and background off-limb emissions.??

\section{Discussion} \label{sec:dis}

Based on the observations, we analyze the triggering process of  activation and possible reasons for the localized brightenings in the filament.
Using the statistic results reported by \citet{2014ApJ...784...50C} and \citet{2017ApJ...835...94O} and the magnetic configuration observed by HMI, which is normal-polarity, we can determine that the magnetic structure of the filament more likely consists of magnetic dips with arcades above.

\par
Before the onset of the activation, the brightened arcade A1 overlying the filament (marked with green dash line in panel (a) in Figure \ref{fig:tri}) rises and reconnects with a nearby small loop (marked with cyan dash line in panel (a) in Figure \ref{fig:tri}), producing brightening at the interface between the small loop and A1 and forming new post-reconnected arcade and small loop (see green and cyan dash lines in panel (b) in Figure \ref{fig:tri}).
This reconnection process destabilizes the filament, %{\bf leading to the outward motions of filament materials along the sight of view and inward motions of the top of the filament (shown as the deeper blue wings of the H$\alpha$ spectral profiles and expansion of red shift region calculated by the position of line core in Figure \ref{fig:dop} and \ref{fig:ras})}
leading to the sinking of filament threads which is shown as the expansion of red shifted region in the dopplergrams (see the black contours in Figure \ref{fig:dop}).
We observe that there exists small misalignment between the localized red shifted regions and localized brightenings (see green contours in panel (d) in Figure \ref{fig:dop}), indicating that the internal reconnections occur among the magnetic field lines with dips supporting the filament threads, rather than among the filament threads.
The sinking of filament threads promotes the internal reconnection between different field lines that support the filament, which might have been braided before the activation, producing the first localized brightening in the filament (see the yellow arrow in panel (c) in Figure \ref{fig:tri}).
The evolution of spectral profiles in the red shifted regions shows deeper blue wings and line centers during the event (see Figure \ref{fig:ras}), suggesting that some filament threads are rising whilst materials are filling in.
During the activation process, the material of filament threads might move along the magnetic dips upwards, which is also manifested as the deeper blue wings of spectral profiles.
Furthermore, destabilization of the filament causes split of the filament threads near the east footpoint (see the blue arrows in panel (c) and (d) in Figure \ref{fig:tri}), which further increases destabilization of the filament.
During this process, the other arcade A3 overlying the filament brightens (see the white dash line in panel (d) in Figure \ref{fig:tri}) and a series of bright points appear near the east footpoint of the filament, which lie on different locations in different AIA EUV passbands (see diamonds in panel (e) in Figure \ref{fig:tri}).
These bright points may resemble the footpoints of arcades overlying the filament which are lightened during the activation process and may promote the internal reconnections between different field lines.
Numerous localized brightenings appear in the filament as a result of a series of internal reconnections.
Parts of these localized brightenings are located at the cross-sections of different filament threads (see the red lines in panel (a) in Figure \ref{fig:dem}), again indicating that these localized brightenings are related to internal reconnections among different field lines that support the filament.

\par
Through the slice-time plots along the filament in AIA 171\,\AA\, passband and SUTRI, we find that a few localized brightenings stay still (see the blue arrows in Figure \ref{fig:st}) while some others propagate along the filament threads with a speed of 50-70 km/s (see the green dash lines in Figure \ref{fig:st}).
%the east end of the filament splits and shifts toward the east end of the brightened loop.
%The newly connected region corresponds to a blue shift region observed in CHASE, and the blue shift enhances after its connection.
%The outflow from the east end destabilizes the filament and the filament begins to erupt and expand.
%During the eruption of the filament, internal reconnections occur and produce series of still and dynamic localized brightenings along the filament.
These stationary localized brightenings appear prior to the moving brightenings, and we believe they are caused by the heating effects of internal reconnections.
Some stationary localized brightenings occur at the same place several times (see the three bottom blue arrows in panel (a) in Figure \ref{fig:st}), which indicates that the internal reconnections occur at the same place repeatedly, and this could play an important role in coronal heating as reported by \citet{2023NatCo..14.2107C}.
For those moving brightenings, their size, multi-thermal nature, speed and density are similar to those found in jets and multi-flux-rope system \citep{2014A&A...567A..11Z,2021RSPSA.47700217S,2018ApJ...857..124A}.
%, though their lifetime, intensity enhancement relative to background is longer and larger,
These moving brightenings are multi-thermal with temperatures in the range of log(T/K) = 5.5-7.3, and their lifetimes, intensity enhancement in relative to the background are several minutes and 0.7-0.8, respectively.
These parameters suggest that they are likely to be the plasmoids or blobs produced in the internal reconnection processes by the tearing-mode instability \citep{2016ApJ...827....4W,2019ApJ...870..113Z}.
%\citet{2016A&A...591A..16P,2016A&A...587A.125P} simulated the evolution of the relax of flux ropes with twisted and braided magnetic field, and compared their appearence in observation.
%In the relaxation process, the filament shows a rotation motion (see the white arrow in Figure \ref{fig:dop} (d)) with no apparent change of shape.
%It is similar to the hypothesis in the simulation of \citet{2022ApJ...927L..14Z} that the transfer between twist and writhe is under the condition with no internal reconnection, as if the internal reconnection occurs, the transfer could not happen.
\citet{2016A&A...591A..16P,2016A&A...587A.125P} compared the behaviour of twisted and braided magnetic field through simulations, and they found that during eruption processes the twisted magnetic fields turn to sigmoid with obvious global rotation while the braided magnetic fields expand significantly without global rotation.
In this event, we obviously observe expansion of the filament during its activation process, which indicates that the magnetic field lines hosting the magnetic dips may be highly braided before the activation.
%We demonstrate that the still localized brightenings are the reconnection regions where internal reconnections caused by braided magnetic arcades take place, and the dynamic localized brightenings are probably the plasmoid produced from the reconnection regions by tearing-mode instability \citep{2016ApJ...827....4W,2019ApJ...870..113Z} as its size, multi-thermal nature, speed and density is similar to those found in jets and multi-flux-rope system \citep{2014A&A...567A..11Z,2021RSPSA.47700217S,2018ApJ...857..124A}, though its lifetime, intensity enhancement relative to background is longer and larger, which is about several minutes and 0.7-0.8, and its multi-thermal range is larger, which is log(T/K) = 5.5-7.3.
%We observed some localized red shift regions near the localized brightenings and this could be the signal of the sink of filament threads.% where greater curvature of magnetic filed is needed to support the filament as a result of expansion of magnetic field.
%In the relaxation process, the filament does not show a sigmoid shape or have an obvious global rotation, but it expands signigicantly.
%This is similar to the relaxation process of the braided field in the simulation of \citet{2016A&A...591A..16P,2016A&A...587A.125P}.

\par
In the later activation process, at 17:00:11\,UT, brightenings at both sides of the filament (see green arrows in panel (g) in Figure \ref{fig:ove}) appear, followed by the unidirectional flow towards eastern footpoint of the filament.
After these bright points and flow, the arcades A2 overlying the filament which can be only observed in AIA 94\,\AA\,passband form.
During the expansion of the filament, some filament threads encounter background magnetic fields, forming current sheet at their interface, which is shown as brightenings at both sides of the filament.
This process is followed by the external reconnections between the filament threads and background magnetic fields, and the unidirectional flow is parts of the outflow of reconnection.
The reconnection process causes the formation of post-reconnected arcades A2 overlying the filaments (see the blue dash lines in panel (c) and red dash lines in panel (l) in Figure \ref{fig:ove}) and heats them to extremely high temperature, which can interpret that the arcades A2 can only be observed in AIA 94\,\AA\,passband.
This indicates that the internal reconnections of the filament promote the occurrence of external reconnections between the filament threads and nearby background magnetic field.
After the activation, the filament cools down in tens of minutes, shown as the gradual disappearance of bright structure in AIA 94\,\AA\,passband, which is in the same order of radiative cooling time scale in the corona \citep{1978ApJ...220..643R,2014LRSP...11....4R}.
%some threads of filament encounter the background fields and reconnect with them, forming hot arcades overlying the filament, which could be only observed in AIA 94\,\AA\. passband.
%The external reconnection is triggered by expansion of the filament, which is caused by the internal reconnection of the filament.
%That is to say, the internal reconnection of the filament prompts the external reconnection of the filament.
%After the eruption, the filament cools down in tens of minutes.

\par
\begin{figure*}
\includegraphics[trim=0cm 0cm 0cm 0cm,width=\textwidth]{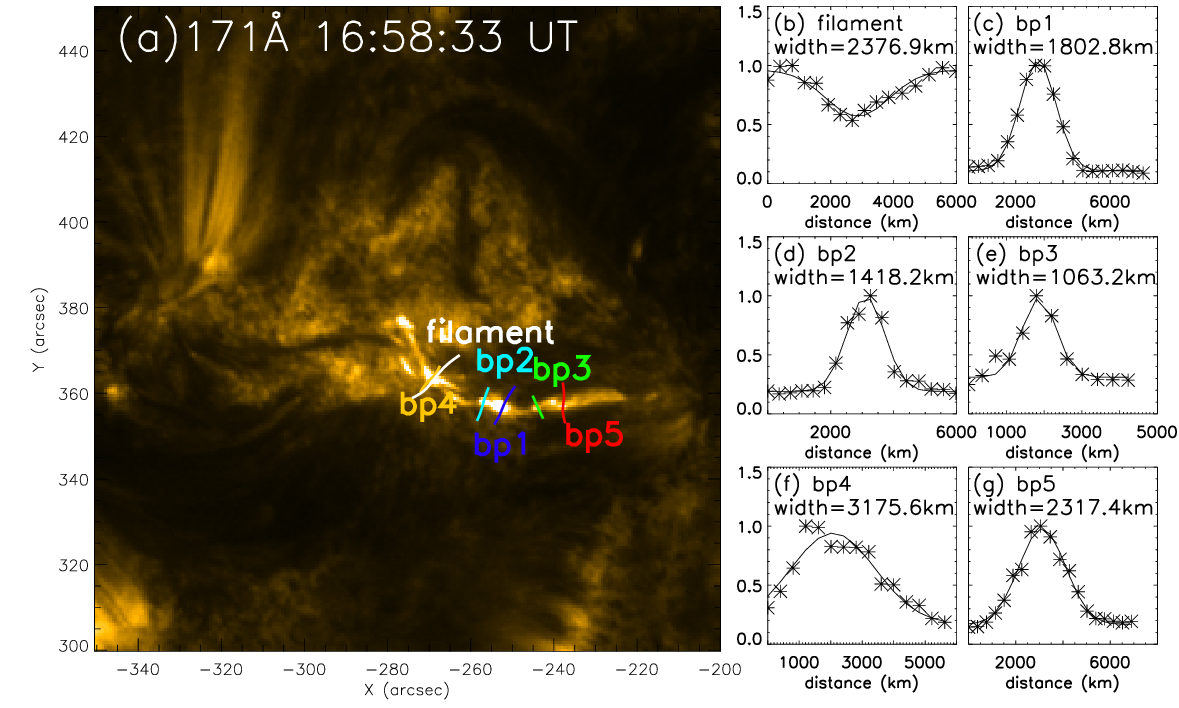}
\caption{Estimation of the width of filament and several localized brightenings.
(a): The image observed in AIA 171\,\AA\, in 16:58:33 UT.
The white line resembles the position where we estimate the width of the filament.
The blue, cyan, green, yellow and red lines resemble the position where we estimate the width of localized brightenings.
(b)-(g): The dotted lines show the intensity distributions along the trajectory in panel (a).
The black lines show the gaussfit results.
\label{fig:wid}}
\end{figure*}

Using the DEM inversion method from \citet{2012ApJ...761...62C}, we obtain the distributions of weight-averaged temperature and density.
Furthermore, we estimate the widths of the filament and several localized brightenings by fitting the intensity distributions of their cross-sections with Gaussian functions (see the width given in Figure \ref{fig:wid}).
The width of the filament is about $2.4 \times 10^8$ cm while those of the localized brightenings are $9.0 \times 10^7$\,cm, $7.1 \times 10^7$\,cm, $5.3 \times 10^7$\,cm, $1.6 \times 10^8$ \,cm and $1.2 \times 10^8$\,cm, respectively.
The initial temperature and density are $3 \times 10^6$\, K and $2 \times 10^{10}$\,cm$^{-3}$, respectively,
while the temperature and density at their peaks are $8 \times 10^6$\,K and $2.8 \times 10^{10} $ cm$^{-3}$, respectively.
As we obtain the parameters of spatial size, temperature and density, we could estimate the thermal energy ($E$) of the active filament threads and localized brightenings via $E = 3N_ek_BTV$, where $N_e$, $k_B$, T and V are the electron number density, Boltzmann constant, temperature and volume, respectively\, \citep{2014ApJ...790L..29T,2020ApJ...899...19C,2021ApJ...918L..20H}.
We could estimate the volume of filament with $V = \frac{1}{4}\pi d^2L$ where d is the width of the filament and L is the length of the filament and the volume of the localized brightenings with $V = \frac{4}{3}\pi r^3$ where r is the radius of brightenings.
The total energy that heats the filament is about $1.77 \times 10^{28} \rm erg$ and the energy that each of the localized brightenings produces is in the order of $10^{25}-10^{27} \rm erg$, which is in the same order of microflares \citep{1991SoPh..133..357H,1998ApJ...501L.213K,2000ApJ...535.1047A,2000ApJ...529..554P,2011SSRv..159..263H}.
%It should be noticed that the heated energy estimated here is about the PCTR, and the total energy in these acitivities might be higher.

%\par
%After the {\bf activation} of the filament, at around 17:30:33\,UT, a series of loops are brightened, and they connect the same positive polarity P1 as the filament and negative polarity N3 next to that of the filament (see the blue 'plus' symbols in Figure \ref{fig:loop} (b)).
%The location of these loops overlaps the arcades A2.
%These loops are most likely formed during the external or internal reconnection processes in the braided magnetic field lines hosting the magnetic dips or arcades overlying the filament that can be transformed into loops through reconnection.
%%The magnetic field of the loop could come from the filament threads, which unbraid during the eruption process, and some of these threads transform into loops after a series of internal reconnections.
%This could provide an explanation to the observational results reported by \citet{2022A&A...667A.166C} that the braiding of loops overlying the chromospheric filamentary are more common.
%We suggest that the loops formed from the filament through reconnections may remain braiding to some extent, which makes it more possible to produce impulsive heating events in their later evolution.

\par
\begin{figure*}
\includegraphics[trim=0cm 0cm 0cm 0cm,width=\textwidth]{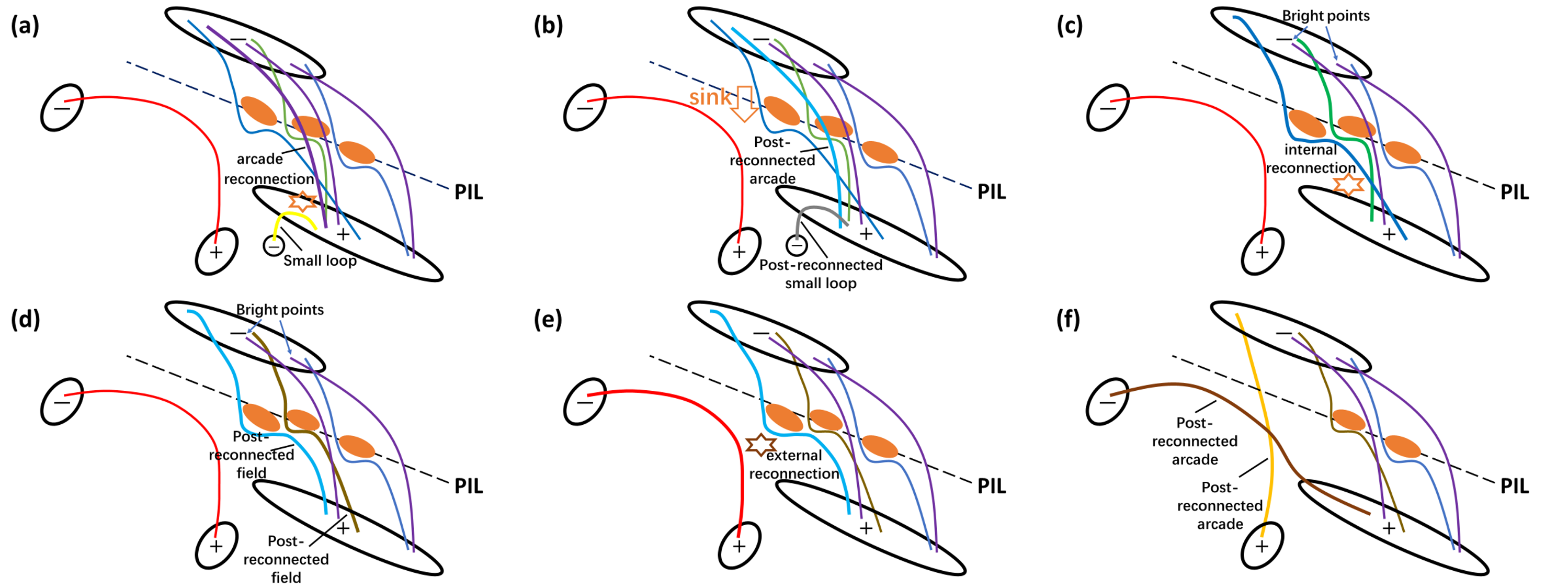}
\caption{A scenario interpreting the reconnection process in this event.
(a): The initial magnetic structures. The green and blue lines resemble the magnetic field lines hosting dips and supporting the filament threads which are shown by orange ovals. The purple lines mark the arcades overlying the filament. The yellow line is the small loop nearby. The red line shows background magnetic field.
(b): The time after reconnection between arcade and nearby loop. The grey and cyan lines mark the post-reconnected loop and arcade respectively.
(c): The internal reconnection between different magnetic field lines hosting dips.
(d): The time after internal reconnection. The cyan and gold lines mark the post-reconnected arcades.
(e): The external reconnection.
(f): The time after external reconnection. The brown line marks the arcade A2 only observed in AIA 94\,\AA\,passband and the yellow line shows another post-reconnected arcade.
\label{fig:car}}
\end{figure*}

According to the observations above, we give a scenario to interpret these processes in Figure \ref{fig:car}.
%Before these activities (Figure \ref{fig:car} (a)), the filament is represented by the black twisted lines connecting anti-polarity.
%The red line resembles the background magnetic field.
%When the filament is destabilized by the outflow at the east end of the filament, it begins to expand and rotate, leading to untwisted or unbraided motion.
%In this process, internal reconnections occur at the cross-section of threads.
%These internal reconnections promote the expansion of the filament and promote the filament threads to encounter the background magnetic field, which causes the external reconnection (Figure \ref{fig:car}).
%The hot loops or arcades form after the external reconnection (see blue line in Figure \ref{fig:car} (d)).
The black ovals marked with plus and minus symbols indicate the positive and negative polarities, and the black dash line represents the position of the PIL.
The red line in panel (a) represents the ambient magnetic field.
Before activities of the filament, the filament threads, marked with orange ovals, are supported by magnetic field lines hosting magnetic dips (green and blue lines) with overlying arcades (purple lines).
The purple and yellow lines in panel (a) mark the arcades overlying the filament and nearby small loop respectively.
Firstly, the arcade reconnects with nearby small loop, which destabilizes the filament, leading to the sinking of the filament threads.
The blue and green lines in panel (b) indicate different magnetic field lines supporting filament threads.
Some of these magnetic lines may be braided before activation of the filament.
The sinking of filament threads promotes the internal reconnections between different magnetic field lines hosting magnetic dips, which then result in brightenings of the footpoints of the arcades overlying the filament (see the blue arrows in panel (c) and (d)).
Series of localized brightenings are produced in these internal reconnection processes with some being stationary and others propagating along post-reconnected arcades.
The stationary brightenings are caused by the heating effects of internal reconnection while the propagating brightenings could be  the blobs or plasmoids caused by tearing-mode instability\,\citep{2016ApJ...827....4W,2019ApJ...870..113Z}.
The internal reconnections are accompanied by the expansion of the filament, and this promotes the external reconnection between the filament threads and background magnetic fields (see panel (e)).
The arcades A2 which can be only observed in AIA 94\,\AA\, passband, marked by brown line in panel (f), form in these external reconnection process.
%Some of the magnetic arcades could be braided.
%After the destabilization of the filament, the filament threads sink and some different braided magnetic arcades reconnect with each other, which produces series of local brightenings.
%In the expansion process of the filament threads, some threads could reconnect with the background magnetic field (see Figure \ref{fig:car} (c)) and form hot arcades (the purple line in Figure \ref{fig:car} (d)).

%\begin{figure*}
%\includegraphics[trim=0cm 0cm 0cm 0cm,width=\textwidth]{fig_hmi.eps}
%\caption{Line-of-sight magnetogram of the east footpoint of the filament (see the yellow dash rectangle in Figure \ref{fig:ove} (c)).
%The red dash rectangle in panel (a) shows the region where we obtain the lightcurve of flux.
%The black arrows represent the horizon motions of the magnetic field.
%(b): The lightcurve of flux in the dash rectangle in panel (a).
%The red dash line indicates the start time of this event.
%\label{fig:hmi}}
%\end{figure*}

\section{Conclusion} \label{sec:con}
In this paper, we report on an eruptive event of a filament with a length of about 70 arcsec observed by SDO, CHASE and SUTRI.
The magnetic geometry of this filament consists of magnetic fields hosting magnetic dips with arcades overlying them.
The activation of the filament can be divided into three stages: 1) the triggering process when the arcades overlying the filament reconnect with ambient loops; 2) internal reconnection process when a series of localized brightenings appear in the filament; 3) external reconnection process when the arcades A2 form.

\par
Before the activation of the filament (stage 1), the arcade A1 overlying the filament brightens and reconnects with a nearby small loop, forming new post-reconnected arcade and loop, which destabilizes the filament and promotes the onset of internal reconnection between different magnetic field lines hosting magnetic dips.
The first localized brightening appears at the middle of filament, located near the localized red-shifted region seen in HIS H$\alpha$ spectral data.
This indicates that this brightening is caused by internal reconnection between the magnetic field lines supporting the filament threads.
The blue wings of spectral profiles in red shifted region deepen during the activation process, and this indicates some threads may rise after the internal reconnection or materials are injected into the filament.
During this process, the filament threads at its east footpoint begin to split, which further destabilizes the filament and promotes the occurrence of internal reconnections.
Meanwhile, numerous bright points appear adjacent to the east footpoint of the filament and their locations differ in different AIA EUV passbands.
These bright points are related to the footpoints of the arcades overlying the magnetic field lines hosting magnetic dips, which may be engaged into the activation process.

\par
In the activation process (stage 2), the filament expands obviously, which is similar to the evolution of braided magnetic filed simulated by \citep{2016A&A...591A..16P,2016A&A...587A.125P}, demonstrating that the magnetic field lines hosting dips may be braided before the activation.
Series of localized brightenings appear in the filament and can be observed in most AIA EUV passbands and SUTRI 465\,\AA\,passband.
These localized brightenings are located at the filament threads and some of them lie at the cross section of different filament threads, which indicates that these localized brightenings are related to the internal reconnections.
Through the slice-time plot along the filament, we find that some localized brightenings are stationary, which are caused by the heating effects of the internal reconnections among magnetic field lines hosting magnetic dips, while others move with speeds ranging from 50 km/s to 70km/s.
These moving brightenings have similar size, multi-thermal nature, speed and density with the blobs or plasmoids in jets and multi-flux-rope system \citep{2014A&A...567A..11Z,2021RSPSA.47700217S,2018ApJ...857..124A}, though there exist some differences in their lifetime, intensity enhancement relative to background and temperature range.
So these moving brightenings are likely to be the blobs or plasmoids produced in the internal reconnection process by tearing-mode instability\,\citep{2016ApJ...827....4W,2019ApJ...870..113Z}.
These localized brightenings show a wide temperature range at log(T/K) = 5.5-7.3, and the weight-averaged temperature is estimated over $8 \times 10^6 \rm K$.
Some filament threads are heated to $8 \times 10^6 \rm K$ from $3 \times 10^6 \rm K$ after series of brightenings.
The total energy released in this heating process is estimated to be $1.77 \times 10^{28} \rm erg$ and the energy each brightening provided is in the range of $10^{25}-10^{27} \rm erg$, which is in the same order of microflares \citep{1991SoPh..133..357H,1998ApJ...501L.213K,2000ApJ...535.1047A,2000ApJ...529..554P,2011SSRv..159..263H}.

% and after that the east end of the filament splits and shifts toward the east end of the loop, which corresponds to the blue shift region.
%The outflow there destabilizes the filament and triggers the eruption of the filament.
%The erupted filament expands with series of localized brightenings.
%These localized brightenings locate on the bright filament threads and some of them locate at the cross section of the bright filament threads.
%Some localized brightenings are still, which could be the internal reconnection regions and could be caused by braid/unbraid motion of the magnetic arcades, while others are dynamic with speeds of 50-70 km/s, which could be the plasmoid from the reconnection sites by tearing-mode instability.

\par
%These brightenings are interpreted as internal reconnection between the filament threads, which could be twisted or braided.
The filament expands obviously during its activation process,  and we observe bright points at both sides of the filament in the stage 3 (see the green arrows in Figure \ref{fig:ove} (g)), followed by an unidirectional flow towards the eastern footpoint of the filament.
These bright points may be the places where the filament threads reconnect with ambient magnetic field while the flow is part of the outflow of external reconnection.
After these external reconnections, arcades A2 overlying the filament which can be only observed in AIA 94\,\AA\, appear.
This demonstrates that the internal reconnection in the activation process can promote the expansion of filament and drive the occurrence of external reconnection.
\textit{Acknowledgement}: We are grateful to the anonymous referee for the constructive comments. This research is supported by National Key R\&D Program of China No. 2021YFA0718600 and National Natural Science Foundation of China (42230203, 42174201, 41974201, 12333009).
CHASE mission is supported by China National Space Administration.
SUTRI is a collaborative project conducted by the National Astronomical Observatories of CAS, Peking University, Tongji University, Xi'an Institute of Optics and Precision Mechanics of CAS and the Innovation Academy for Microsatellites of CAS.
The AIA and HMI data are used by courtesy of NASA/SDO, the AIA and HMI teams and JSOC.
%\end{acknowledgments}






\bibliography{reference}{}
\bibliographystyle{aasjournal}



\end{document}